\documentclass[prl,aps,twocolumn,amsmath,amssymb,floatfix,superscriptaddress]{revtex4}

\usepackage[dvips]{graphics}
\usepackage{color}
\usepackage{soul}
\usepackage[colorlinks=true, citecolor=blue, urlcolor=blue]{hyperref}

\date{\today}

\begin{document}

\title{Estimating critical disorder strength of an AAH system via thermal response}

\author{Sanchita Nandi}

\affiliation{Physics and Applied Mathematics Unit, Indian Statistical Institute, 203 Barrackpore Trunk Road, Kolkata-700 108, India}

\author{Santanu K. Maiti}

\email{santanu.maiti@isical.ac.in}

\affiliation{Physics and Applied Mathematics Unit, Indian Statistical Institute, 203 Barrackpore Trunk Road, Kolkata-700 108, India}

\begin{abstract}

The Aubry-Andr\'e-Harper (AAH) model provides a paradigmatic platform to study localization phenomena, where a transition from a 
conducting to an insulating phase occurs at a critical disorder strength equal to twice the nearest-neighbor hopping amplitude, $2t$. 
While this transition has been explored extensively through various probes, here we propose a fundamentally new approach based on 
thermal response. Specifically, we demonstrate that the electronic specific heat exhibits a distinctive signature across the transition, 
directly reflecting the anomalous evolution of the density of states with disorder strength. Our findings, supported by numerical analysis, 
establish specific heat measurement as a powerful and sensitive probe for detecting the critical point. This thermal characterization 
opens a new avenue for exploring disorder-induced phase transitions and paves the way for experimental realization in engineered 
quantum systems.

\end{abstract}

\maketitle

Disorder-induced quantum phase transitions in low-dimensional systems have remained a central topic in condensed matter physics since the seminal work of Anderson on localization~\cite{lc1}. In low-dimensional lattices with random disorder, any finite amount of disorder is sufficient to localize all electronic states~\cite{lc1,lc2,lc3,lc4,lc5,lc6}, thereby driving a phase transition from a conducting to an 
insulating state. This implies that the critical disorder strength is effectively zero in such systems~\cite{lc1,lc2,lc3,lc4,lc5,lc6}, 
highlighting the extreme sensitivity of low-dimensional transport to randomness.

A particularly intriguing departure from this behavior occurs where disorder is introduced in a quasiperiodic~\cite{qc1,qc2,qc3} rather 
than a random manner. The Aubry-Andr\'e-Harper (AAH) model serves as a canonical example in this class~\cite{abr1,abr2,abr3,abr4}. 
In this model, 
the on-site potentials are modulated quasiperiodically, typically as $\epsilon_n=W \cos(2 \pi \beta n + \phi)$, where $W$ is the modulation strength, $\beta$ is an irrational number, and $\phi$ is the phase factor. Remarkably, unlike random disorder, this system exhibits a sharp
localization-delocalization transition at a finite critical modulation strength $W_c=2t$, where $t$ is the nearest-neighbor hopping 
amplitude~\cite{abr2,abr3,abr4}. For $W<2t$, all eigenstates are extended, while for $W>2t$, all states are localized, making it an 
ideal platform for studying deterministic disorder-induced transitions in one dimension.

This transition has been extensively investigated using a wide range of theoretical diagnostics. These include the inverse participation ratio (IPR)~\cite{abr5,ipr1,ipr2,ipr3}, which quantifies the spatial localization of eigenstates, as well as statistical measures such as level spacing distributions. Transport-based quantities like conductance and Lyapunov exponents~\cite{lp1,lp2}, and dynamical measures~\cite{dyn1} such as wavepacket spreading and entanglement entropy~\cite{enp1,enp2,enp3} have also been used to detect and characterize the transition. Despite their effectiveness in revealing the localization physics, these probes are fundamentally tied to eigenvalues and eigenstates, and are often inaccessible in direct experimental measurements.

Experimentally, the detection of the AAH transition has primarily been confined to synthetic systems, such as ultracold atoms 
in optical lattices~\cite{oplt1,oplt2} and photonic crystals~\cite{phl1,phl2,phl3}, where the requisite quasiperiodic potential landscapes 
can be engineered with high precision. These platforms have enabled remarkable demonstrations of localization transitions analogous to 
those predicted by the AAH model. More recently, there have been exploratory studies involving mechanical metamaterials and 
nano-electromechanical resonator arrays~\cite{meta,meca}, expanding the experimental reach of AAH-type physics. However, despite these 
advances, {\em the realization and detection of the AAH transition in purely electronic 1D systems remains severely limited.} 
The quantum Hall system~\cite{qh1,qh2,qh3}, involving 2D electron gases under strong magnetic fields, stands as one of the rare examples 
where AAH-like behavior emerges in electronic settings. Yet this system is fundamentally two-dimensional and magnetic-field-dependent, 
differing in essential ways from the 1D AAH scenario.

In light of this gap, the present Letter proposes a new and experimentally feasible route to detect 
\begin{figure}[ht]
{\centering\resizebox*{8cm}{0.7cm}{\includegraphics{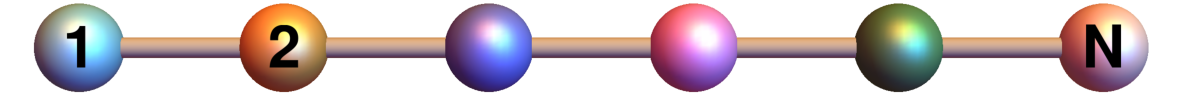}}\par}
\caption{(Color online). Schematic of a 1D AAH chain with $N$ lattice sites. Different colored balls are used to represent the variation 
in site energies across the lattice.}
\label{figchain}
\end{figure}
the AAH transition in one-dimensional electronic systems. Specifically, we report, for the first time, that the transition can be probed 
through the measurement of the electronic specific heat (ESH), a well-established quantity~\cite{sh1,sh2,sh3,sh4} that is routinely 
measured in laboratory settings~\cite{spexp1,spexp2,spexp3}. Using extensive numerical calculations, we demonstrate that the ESH exhibits 
a distinct and robust signature across the AAH transition point. This response is directly linked to the anomalous restructuring of the 
electronic density of states (DOS) as the modulation strength $W$ crosses the critical value.

The strength of our approach lies in its direct experimental relevance, in contrast to existing theoretical probes that rely on spectral 
or state-level diagnostics. Moreover, the method is general and versatile, holding across a broad range of physically relevant parameters, 
and does not require any exotic experimental apparatus. Our findings thus open a promising direction for the experimental realization of 
quasiperiodic localization phenomena in low-dimensional electronic systems, potentially enabling new classes of material diagnostics 
based on thermal measurements.

We consider a one-dimensional tight-binding chain with site energies modulated according to the AAH model. 
A schematic representation of the system is shown in Fig.~\ref{figchain}. The Hamiltonian of the system is given by~\cite{abr4,abr5,phl1}
\begin{equation}
H = \sum_n \left[ \epsilon_n c_n^\dagger c_n + t \left(c_n^\dagger c_{n+1} + c_{n+1}^\dagger c_n \right) \right],
\label{hamil}
\end{equation}
where $c_n^\dagger$ and $c_n$ are the fermionic creation and annihilation operators at site $n$, respectively, and $t$ is the 
nearest-neighbor hopping amplitude. The on-site energies $\epsilon_n$ are modulated as~\cite{abr4,abr5,phl1}
\begin{equation}
\epsilon_n = W \cos(2\pi b n),
\label{energy}
\end{equation}
with $W$ being the modulation strength. We choose $b=(1+\sqrt{5})/2$, the golden mean, as a representative irrational number commonly used 
in the literature~\cite{abr4,abr5,phl1}, although the physical behavior remains qualitatively similar for other incommensurate choices. 

To probe the thermal response of the system, we focus on the electronic specific heat $C_v$, which is directly connected to the density of
states. The density of states, $\rho(E)$, is calculated using the retarded Green's function formalism~\cite{gf1}
\begin{equation}
\rho(E) = -\frac{1}{N\pi} \text{Im} \left[ \text{Tr}\left(G(E)\right) \right],
\label{dos}
\end{equation}
where $G(E)$ is the Green's function defined as~\cite{gf1,gf2}
\begin{equation}
G(E) = \left( E - H + i\eta \right)^{-1},
\label{green}
\end{equation}
with $\eta\rightarrow 0^{+}$. The eigenenergies $E_i$ are obtained via direct diagonalization of the Hamiltonian.

The electronic specific heat $C_v$ is derived from the temperature derivative of the system's average 
energy $\bar{U}$~\cite{sh1,sh2,sh3,sh4}
\begin{equation}
C_v = \frac{d \bar{U}}{dT},
\label{cv}
\end{equation}
where
\begin{equation}
\bar{U} = \sum_{i=1}^N (E_i - \mu) f(E_i).
\label{avgenergy}
\end{equation}
Here, $E_i$ are the eigenenergies obtained via direct diagonalization of the Hamiltonian, $\mu$ denotes the electrochemical potential, and 
$f(E_i)$ is the Fermi-Dirac distribution function, given by 
\begin{equation}
f(E_i) = \frac{1}{1 + \exp\left[(E_i - \mu)/(k_B T)\right]},
\label{fermi}
\end{equation}
where $k_B$ is the Boltzmann constant and $T$ is the absolute temperature. This thermodynamic formulation allows us to capture the collective response of all accessible electronic states, thereby enabling a direct assessment of ESH encoded in the DOS.

We now present and analyze our results. The central objective of this Letter is to determine the critical disorder strength $W_c$ 
characterizing the AAH transition, from the measurement of electronic specific heat. Before delving into the results, we outline the
parameter values and conventions adopted throughout the analysis. All energy values are expressed in electron-volt (eV), temperature 
is given in Kelvin (K), and the Boltzmann constant is taken as $k_B=8.6173 \times 10^{-5}\,$eV/K. The ESH is computed by fixing 
the total number of electrons, $N_e$, and the electrochemical potential $\mu$ is determined self-consistently for each case. In our
analysis, ESHs are calculated for the half-filled band case, without considering spin degree of freedom. Unless otherwise specified, 
the system size is set to $N=200$ lattice sites, and the nearest-neighbor hopping amplitude is chosen as $t=0.2\,$eV. The choice of a 
relatively small hopping strength is intentional: due to the $10^{-5}$ factor in $k_B$, smaller values of $t$ ensures that the ESH 
exhibits discernible features within experimentally accessible temperature ranges. A more detailed justification for such a choice 
is available in Ref.~\cite{sh4}.

\begin{figure}[ht]
{\centering\resizebox*{7cm}{5cm}{\includegraphics{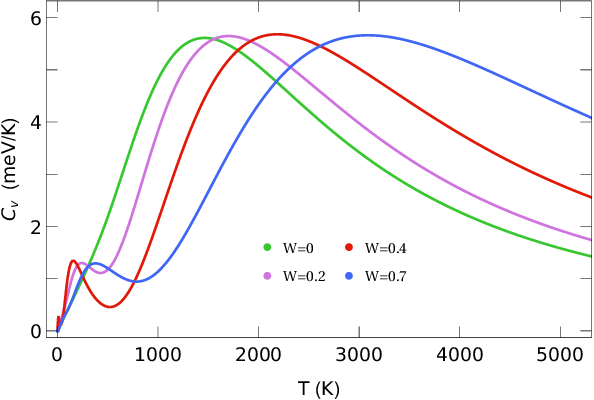}}\par}
\caption{(Color online). Variation of electronic specific heat as a function of temperature for a 1D AAH chain at various disorder 
strengths $W$. The result for a perfect ($W=0$) chain is also included for comparison.} 
\label{cvtemp}
\end{figure}

\begin{figure}[ht]
{\centering\resizebox*{8cm}{7cm}{\includegraphics{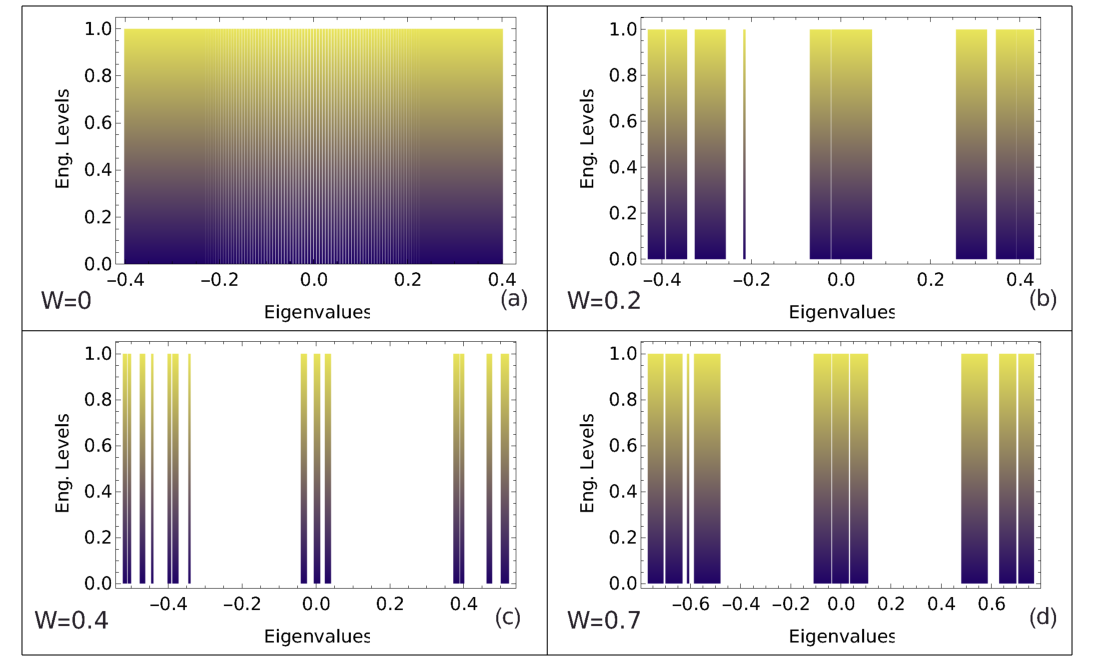}}\par}
\caption{(Color online). Energy eigenspectra of the 1D AAH chain for different disorder strengths, using the same parameter set as in 
Fig.~\ref{cvtemp}, where (a) $W=0$, (b) $W=0.2$, (c) $W=0.4$, and (d) $W=0.7$. Each eigenvalue is indicated by a vertical line of unit 
height for visual clarity.}
\label{englevels}
\end{figure}

We begin with Fig.~\ref{cvtemp}, which shows the temperature dependence of ESH over a broad range of temperatures for a 1D AAH chain at representative disorder strengths. For comparison, we also include the result for a perfectly ordered chain ($W=0$).
Three representative disorder strengths are considered: $W=0.2$ (below $W_c$), $W=0.4$ (at $W_c$), and $W=0.7$ (above $W_c$). 
The ESH exhibits several notable features. For the clean system, $C_v$ increases with temperature, reaches a maximum, and then 
decreases -- a typical thermal response~\cite{sh3,sh4}. In contrast, the disordered chains exhibit a distinctive `valley' in $C_v$ 
at low temperatures, beyond which the ESH recovers a similar qualitative trend as the clean system. This valley is most pronounced 
at the critical disorder strength $W=W_c$. These behaviors are intimately connected to the underlying energy eigenspectrum, as 
illustrated in Fig.~\ref{englevels}. 

For the perfect chain, the energy levels are nearly uniformly spaced throughout the allowed energy window, except near the band edges. 
At absolute zero temperature, energy levels are occupied up to the chemical potential. As temperature increases, thermally accessible 
states around $\mu$ begin to contribute, raising the internal energy $\bar{U}$, and thus $C_v$. Initially, the rate of increase in 
$\bar{U}$ with temperature is substantial, but as temperature rises further and more states become thermally accessible, this rate 
saturates, resulting in a peak followed by a gradual decline in $C_v$. 

In contrast, for the AAH chain, the energy levels are fragmented into three sub-bands, separated by significant gaps. For our chosen 
filling, the chemical potential $\mu$ lies within the central sub-band. At low temperatures, the behavior of $\bar{U}$ resembles 
that of the clean system; however, the presence of a large inter-band gap inhibits the thermal excitation of electrons across energy
levels. This suppression leads to 
a drop in $C_v$, forming the observed valley. Once the temperature becomes large enough to bridge the gap, $\bar{U}$ and thus $C_v$ 
increases again, peaking and then decreasing, as in the clean limit. The valley becomes particularly prominent at $W=W_c$, where 
the sub-bands are maximally narrowed and the energy gaps between them are most significant. This unique feature provides a clear 
signature of the critical point.

\begin{figure}[ht]
{\centering\resizebox*{7cm}{5cm}{\includegraphics{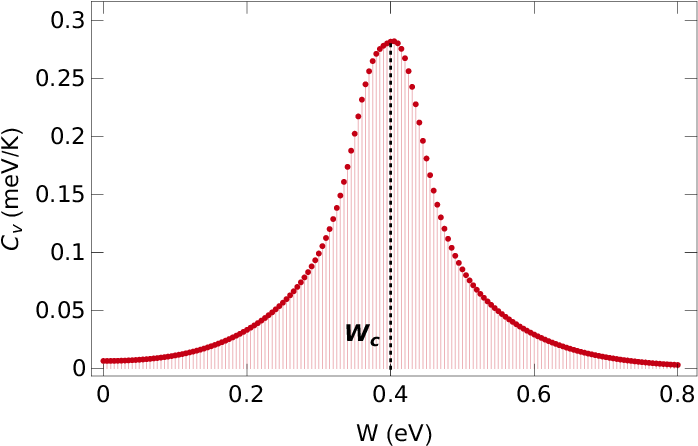}}\par}
\caption{(Color online). ESH as a function of disorder strength $W$ at a fixed temperature of $10\,$K. The ESH exhibits a maximum at the critical disorder strength $W=W_c$.}
\label{cvdis}
\end{figure}

\begin{figure}[ht]
{\centering\resizebox*{7.3cm}{5cm}{\includegraphics{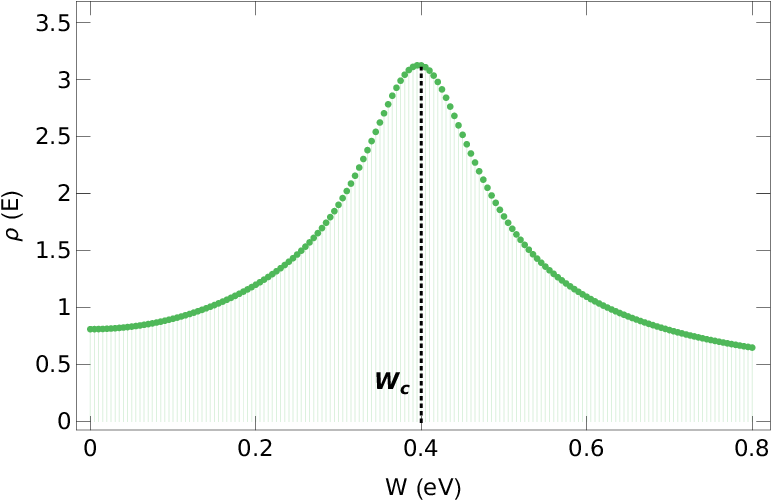}}\par}
\caption{(Color online). Density of states as function of $W$ for the half-filled band case. A pronounced maximum appears at $W=W_c$.}
\label{rhovsW}
\end{figure}

To further validate this observation, Fig.~\ref{cvdis} shows the variation of $C_v$ as a function of disorder strength $W$ at a fixed 
temperature of $10\,$K. A clear peak appears in the $C_v$-$W$ profile at $W_c=0.4$. This trend rises from the systematic narrowing of 
the energy sub-bands with increasing $W$. As the sub-bands become narrower, more energy levels cluster within a smaller energy window, 
enhancing the density of states (DOS) around the chemical potential. This trend is quantitatively supported by the DOS analysis shown 
in Fig.~\ref{rhovsW}, where $\rho$ is calculated at $\mu$ for the half-filled band case. The higher availability of states within the 
energy window $k_BT$ leads to enhanced thermal excitation and, 
consequently, a higher $C_v$. This explains the observed maximum in $C_v$ at the critical disorder strength $W_c$.

\begin{figure*}[ht]
{\centering\resizebox*{5.5cm}{5.5cm}{\includegraphics{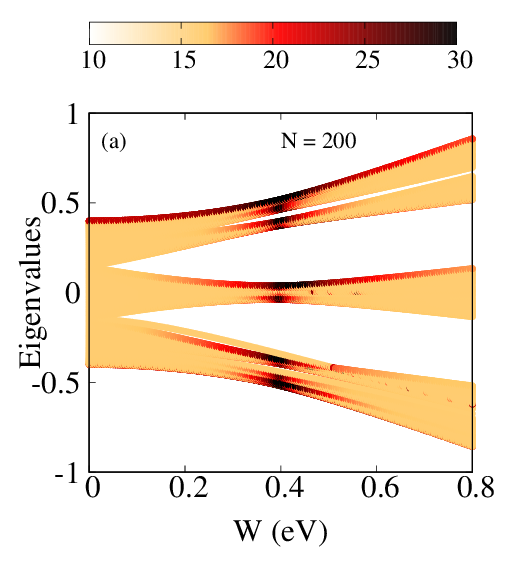}}
\resizebox*{5.5cm}{5.5cm}{\includegraphics{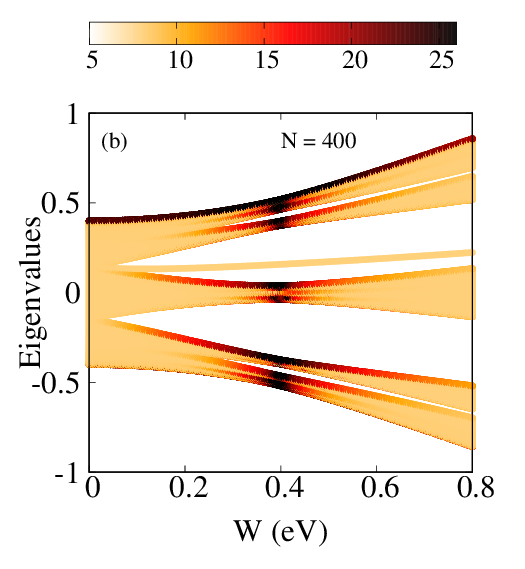}}
\resizebox*{5.5cm}{5.5cm}{\includegraphics{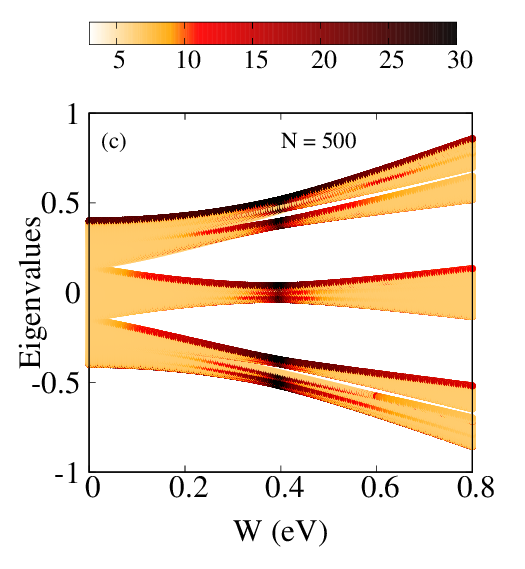}}\par}
\caption{(Color online). Density plots of DOS as functions of energy eigenvalues and disorder strength for three different chain lengths, where (a), (b), and (c) correspond to $N=200$, $400$, and $500$, respectively. For each system size, the DOS is calculated at all eigenvalues for varying $W$. A clear maximum in the DOS consistently appears at the critical point $W=W_c$, irrespective of energy.}
\label{fig6}
\end{figure*}

To verify the robustness of this peak across different system sizes and energy scales, we extend our analysis in Fig.~\ref{fig6}, where we
present density plots of the DOS as functions of energy and disorder strength for three different chain lengths: $N=200$, $400$, and $500$.
Across all system sizes, a pronounced enhancement in the DOS is consistently observed at $W=W_c$, irrespective of the energy. This implies 
that the associated peak in $C_v$ is not restricted to a particular size but represents a robust feature of the system. Moreover, since 
the DOS enhancement at
$W_c$ is evident across the entire energy spectrum, the associated peak in $C_v$ should appear regardless of the filling fraction. 
This highlights the generality and reliability of ESH as a diagnostic tool for detecting the AAH transition in purely electronic systems.

In conclusion, we have demonstrated that the electronic specific heat serves as a sensitive and experimentally viable probe for detecting 
the critical disorder strength in the Aubry-Andr\'e-Harper model. A pronounced thermal signature, peaking at the transition point, emerges 
from the restructuring of the density of states across the localization boundary. Unlike conventional spectral or transport-based 
diagnostics, this approach offers a practical pathway to identify quasiperiodic phase transitions using standard calorimetric 
techniques. Our results open up new opportunities for exploring localization in low-dimensional systems through thermal response.

\end{document}